\documentclass[preprintnumbers,amsmath,amssymb,10pt, showpacs, twocolumn]{revtex4}

\usepackage{graphicx}
\usepackage{dcolumn}
\usepackage{bm}

\begin{document}

\title{Remote Preparation of the Two-Particle State}

\author{YAN Feng-Li, ZHANG Guo-Hua}

\affiliation {College of Physics Science and Information
Engineering, Hebei Normal
University, Shijiazhuang 050016, China\\
}

\begin{abstract}
We present a  scheme of  remote preparation of the two-particle
state
 by using two Einstein-Podolsky-Rosen pairs or two partial entangled
two-particle states as the quantum channel. The probability of the successful  remote state preparation is
obtained.
\end{abstract}

\pacs{03.65.Ta, 03.67.Hk, 03.67.Lx}

 \maketitle

Quantum entanglement is a valuable resource  for the implementation
of quantum computation and quantum communication protocols, like
quantum teleportation [1,2], quantum key distribution [3,4], quantum
secure direct communication [5], dense coding [6-8], quantum
computation [9], remote state preparation [10-15] and so on. Quantum
teleportation is regarded as one of the most profound results of
quantum information theory. In the original quantum teleportation
protocol of Bennett et al. [1], it was  showed that an unknown state
of a qubit can be perfectly transported from a sender Alice to a
remote receiver Bob with the aid of long-range
Einstein-Podolsky-Rosen  correlation and transmission of two bits of
classical information without transmission of the carrier of the
quantum state. They have also generalized the protocol for an
unknown qubit by using a maximally entangled state in $d\times d$
dimensional Hilbert space and sending $2{\rm log}_2 d$ bits of
classical information.

Another important application of quantum entanglement is remote
state preparation, where two spatially distant people Alice and Bob
can prepare an
 quantum state known to Alice but unknown to Bob,
 with the aid of previously shared
 quantum entanglement and the classical communication.
 Recently, Pati [10]  presented a protocol of remotely preparing
a special ensemble of states.  Lo [11] showed that  remote state preparation requires less classical
communication than teleportation for the special ensembles of states,  but for general states the classical
communication cost of teleportation  would be equal to that of remote state preparation. Bennett et al. [12]
studied the trade off between entanglement cost and classical communication cost in remote state preparation.
Since then, some people have investigated various theoretical protocols about generalization of remote state
preparation.   Zhan [13]  gave a scheme for preparing remotely a three-particle GHZ state.  Huang et al. [14]
 put forward a protocol for preparing remotely the multipartite pure state.  A scheme for preparing remotely
a two-particle entangled state via two pairs of entangled particles was presented by  Liu et al. [15].

In this paper, we propose a scheme of preparing remotely the two-particle state.  Two Einstein-Podolsky-Rosen
pairs  and two partial entangled two-particle states as the quantum channel are considered, respectively. The
 probability of the successful remote state preparation is  calculated.

Let us first  begin our remote state preparation  with  two
Einstein-Podolsky-Rosen pairs as the quantum channel. We suppose
that a sender Alice wants to help a remote receiver Bob to  prepare
a two-particle state in the following formation
\begin{equation}
|\phi\rangle=\alpha|00\rangle+\beta|01\rangle+\gamma|10\rangle+\delta|11\rangle,
\end{equation}
where $\alpha$ and $\gamma$ are real numbers, $\beta$ and $\delta$
are complex numbers and $\beta^*\delta$ is real,
$|\alpha|^2+|\beta|^2+|\gamma|^2+|\delta|^2=1$. We suppose that
Alice knows $\alpha$, $\beta$, $\gamma$ and $\delta$ completely, but
Bob does not know them at all. We also assume that the quantum
channel shared by Alice and Bob is composed of two
Einstein-Podolsky-Rosen pairs
 \begin{equation}
|\psi\rangle_{12}=\frac {1}{\sqrt 2}(|00\rangle+|11\rangle)_{12},
 \end{equation}
\begin{equation}
|\psi\rangle_{34}=\frac {1}{\sqrt 2}(|00\rangle+|11\rangle)_{34},
 \end{equation}
where particles 1 and 3 belong to Alice while Bob has particles 2
and 4. In order to help Bob to remotely prepare a two-particle state
stated in Eq.(1) on the particles 2 and 4, Alice must make a
measurement on her two particles 1 and 3. The measurement basis
chosen by Alice is a set of mutually orthogonal basis vectors
$\{|\varphi\rangle_{13},|\varphi_\perp\rangle_{13},|\psi\rangle_{13},|\psi_\perp\rangle_{13}\}$,
where
\begin{eqnarray}
&&\nonumber|\varphi\rangle_{13}=(\alpha|00\rangle+\beta|01\rangle+\gamma|10\rangle+\delta|11\rangle)_{13},\\\nonumber
&&|\varphi_\perp\rangle_{13}=(-\delta^*|00\rangle+\gamma|01\rangle-\beta^*|10\rangle+\alpha|11\rangle)_{13},\\\nonumber
&&|\psi\rangle_{13}=(\gamma|00\rangle+\delta|01\rangle-\alpha|10\rangle-\beta|11\rangle)_{13},\\\nonumber
&&|\psi_\perp\rangle_{13}=(\beta^*|00\rangle-\alpha|01\rangle-\delta^*|10\rangle+\gamma|11\rangle)_{13}.\\\end{eqnarray}
Here
$\{|00\rangle_{13},|01\rangle_{13},|10\rangle_{13},|11\rangle_{13}
\}$ is the computation basis. Then we have
\begin{eqnarray}
&&\nonumber|\psi\rangle_{12}\otimes|\psi\rangle_{34}\\\nonumber
&=&\frac
{1}{2}[|\varphi\rangle_{13}(\alpha|00\rangle+\beta^*|01\rangle+\gamma|10\rangle+\delta^*|11\rangle)_{24}\\\nonumber
&&+|\varphi_\perp\rangle_{13}(-\delta|00\rangle+\gamma|01\rangle-\beta|10\rangle+\alpha|11\rangle)_{24}\\\nonumber
&&+|\psi\rangle_{13}(\gamma|00\rangle+\delta^*|01\rangle-\alpha|10\rangle-\beta^*|11\rangle)_{24}\\\nonumber
&&+|\psi_\perp\rangle_{13}(\beta|00\rangle-\alpha|01\rangle-\delta|10\rangle+\gamma|11\rangle)_{24}].\\
\end{eqnarray}

Thus if  Alice performs a measurement in the basis
$\{|\varphi\rangle_{13},|\varphi_\perp\rangle_{13},|\psi\rangle_{13},|\psi_\perp\rangle_{13}\}$
on her two particles 1 and 3, then each outcome will occur with the
equal probability $\frac {1}{4}$. If Alice's measurement result is
$|\varphi_\perp\rangle$,  then particles 2 and 4 will collapse into
the state
\begin{equation}
(-\delta|00\rangle+\gamma|01\rangle-\beta|10\rangle+\alpha|11\rangle)_{24}.
\end{equation}
After that  if Alice communicates to  Bob of her actual measurement
outcome via a classical channel, then Bob will be able to  apply the
following unitary transformation
\begin{equation}
U=(|0\rangle\langle 1|+|1\rangle \langle
0|)_2\otimes(|0\rangle\langle 1|-|1\rangle \langle 0|)_4
\end{equation}
on his particles 2 and 4. The resulting state of  Bob's particles
will be  the original state $|\phi\rangle$.

Likewise if the actual result of Alice's measurement  is
$|\psi_\perp\rangle_{13}$, then Bob gets the state
\begin{equation}
(\beta|00\rangle-\alpha|01\rangle-\delta|10\rangle+\gamma|11\rangle)_{24}.
\end{equation}
When Bob received the  classical information of the actual
measurement result sent by Alice, he can perform an appropriate
operation
\begin{equation}
U'=(|0\rangle\langle 0|-|1\rangle \langle
1|)_2\otimes(|0\rangle\langle 1|-|1\rangle \langle 0|)_4
\end{equation}
on his particles 2 and 4 to obtain  the state  $|\phi\rangle$. So
when these two measurement outcomes happen, Alice can help  Bob to
remotely prepare the two-particle state
 $|\phi\rangle$.

 However, when the  measurement outcome is  $|\varphi\rangle_{13}$  ($|\psi\rangle_{13}$),
the remote state preparation can not be successful,  as the state of
the particles 2 and 4 will be
$(\alpha|00\rangle+\beta^*|01\rangle+\gamma|10\rangle+\delta^*|11\rangle)_{24}$
(
$(\gamma|00\rangle+\delta^*|01\rangle-\alpha|10\rangle-\beta^*|11\rangle)_{24}$).
Because Bob does not know the coefficients $\alpha$, $\beta$,
$\gamma$ and $\delta$ at all,  he cannot transform either the state
$(\alpha|00\rangle+\beta^*|01\rangle+\gamma|10\rangle+\delta^*|11\rangle)_{24}$
or
$(\gamma|00\rangle+\delta^*|01\rangle-\alpha|10\rangle-\beta^*|11\rangle)_{24}$
into the state $|\phi\rangle$. But,  if  $\alpha$, $\beta$, $\gamma$
and $\delta$ are  real numbers, the situation would be changed. When
Alice's measurement result  $|\varphi\rangle_{13}$ or
$|\psi\rangle_{13}$ occurs,  it is not difficult for Bob to  prepare
the two-particle state $|\phi\rangle$ by performing the  suitable
unitary operation determined by the outcome of Alice's measurement.
Here we omit the concrete steps.

Now, we present a scheme for preparing remotely a two-particle state
 via two non-maximally entangled states. Suppose
that  Alice still wishes to help Bob to prepare remotely the state
$|\phi\rangle$ in Eq.(1), but  the two entangled states shared by
Alice and Bob are two non-maximally entangled states
\begin{equation}
|\psi\rangle_{12}=(a|00\rangle+b|11\rangle)_{12},
\end{equation}
\begin{equation}
|\psi\rangle_{34}=(c|00\rangle+d|11\rangle)_{34},
\end{equation}
where the parameters  $a$, $b$, $c$ and $d$  are real numbers,
$|a|^2+|b|^2=1$, $|c|^2+|d|^2=1$, and $|a|\leq |b|,$ $|c|\leq |d|$.
We also assume that the particles 1 and 3 belong to Alice while Bob
has particles 2 and 4. Since Alice knows the parameters $\alpha$,
$\beta$, $\gamma$  and $\delta$ exactly, she can perform a
 measurement on particles 1 and 3 in the basis
$\{|\varphi\rangle_{13},|\varphi_\perp\rangle_{13},|\psi\rangle_{13},|\psi_\perp\rangle_{13}\}$.
A simple algebraic rearrangement of the expression
$|\psi\rangle_{12} \otimes |\psi\rangle_{34}$ in terms of the states
$|\varphi\rangle_{13},$
 $|\varphi_\perp\rangle_{13},$ $|\psi\rangle_{13},$ $|\psi_\perp\rangle_{13}$
yields
\begin{eqnarray}
&&\nonumber|\psi\rangle_{12}\otimes|\psi\rangle_{34}\\\nonumber
&=&|\varphi\rangle_{13}(ac\alpha|00\rangle+ad\beta^*|01\rangle+bc\gamma|10\rangle+bd\delta^*|11\rangle)_{24}\\\nonumber
&&+|\varphi_\perp\rangle_{13}(-ac\delta|00\rangle+ad\gamma|01\rangle-bc\beta|10\rangle+bd\alpha|11\rangle)_{24}\\\nonumber
&&+|\psi\rangle_{13}(ac\gamma|00\rangle+ad\delta^*|01\rangle-bc\alpha|10\rangle-bd\beta^*|11\rangle)_{24}\\\nonumber
&&+|\psi_\perp\rangle_{13}(ac\beta|00\rangle-ad\alpha|01\rangle-bc\delta|10\rangle+bd\gamma|11\rangle)_{24}.\\
\end{eqnarray}
Therefore,  if the actual result of Alice's measurement on the two
particles 1 and 3 is $|\varphi_\perp\rangle_{13}$ with the
probability $|bd\alpha|^2+|bc\beta|^2+|ad\gamma|^2+|ac\delta|^2$
then the state of particles 2 and 4  will be
\begin{equation}
|\varphi\rangle_{24}=\frac
{(-ac\delta|00\rangle+ad\gamma|01\rangle-bc\beta|10\rangle+bd\alpha|11\rangle)_{24}}{\sqrt
{|bd\alpha|^2+|bc\beta|^2+|ad\gamma|^2+|ac\delta|^2}}.
\end{equation}
When Bob is informed the actual measurement outcome
$|\varphi_\perp\rangle_{13}$ by Alice via a classical channel, he
can get   the original state described in Eq.(1) with certain
probability.

Firstly, Bob operates a unitary operation
\begin{equation}
U_1=(|0\rangle\langle 1|+|1\rangle\langle
0|)_2\otimes(|0\rangle\langle 1|-|1\rangle\langle 0|)_4
\end{equation}
on particles 2 and 4. Obviously $U_1$ will transform the state
$|\varphi\rangle_{24}$ into
\begin{equation}
|\varphi'\rangle_{24}=\frac
{(bd\alpha|00\rangle+bc\beta|01\rangle+ad\gamma|10\rangle+ac\delta|11\rangle)_{24}}{\sqrt
{|bd\alpha|^2+|bc\beta|^2+|ad\gamma|^2+|ac\delta|^2}}.
\end{equation}
Secondly, Bob introduces an auxiliary two-level particle $a$ with
the initial state $|0\rangle_a$  and performs a collective unitary
transformation
\begin{eqnarray} U_2=\left (\begin{array}{cccccccc}
\frac
{ac}{bd}&A&0&0& 0&0&0&0\\
A&-\frac {ac}{bd}&0&0&0&0&0&0\\
0&0&\frac {a}{b}&B&0&0&0&0\\
0&0&B&-\frac {a}{b}&0&0&0&0\\
0&0&0&0&\frac
{c}{d}&C&0&0\\
0&0&0&0&C&-\frac {c}{d}&0&0\\
0&0&0&0&0&0&1&0\\
0&0&0&0&0&0&0&-1\\
\end{array}\right)\end{eqnarray}
 on particles 2, 4 and $a$ under the basis $\{
|000\rangle_{24a},|001\rangle_{24a},|010\rangle_{24a},
|011\rangle_{24a},|100\rangle_{24a},|101\rangle_{24a},$
$|110\rangle_{24a},|111\rangle_{24a}\},$ where
\begin{equation}\nonumber\begin{array}{ccc}
A=\sqrt {1-(\frac {ac}{bd})^2},& B=\sqrt {1-(\frac
{a}{b})^2},&C=\sqrt {1-(\frac {c}{d})^2}.\end{array}\end{equation}

Since it has been  assumed  that $|a|\leq |b|$ and $|c|\leq |d|$, so
one has $|ac|^2\leq |bd|^2$. The unitary transformation $U_2$ will
transform $|\varphi'\rangle_{24}|0\rangle_a$ into
\begin{eqnarray}
\nonumber|\varphi''\rangle_{24a} &=&\frac {1}{\sqrt
{|bd\alpha|^2+|bc\beta|^2+|ad\gamma|^2+|ac\delta|^2}}\\\nonumber
&&{[ac(\alpha|00\rangle+\beta|01\rangle+\gamma|10\rangle+\delta|11\rangle)_{24}}|0\rangle_a\\\nonumber
&&+(\alpha\sqrt {(bd)^2-(ac)^2}|00\rangle+c\beta\sqrt
{b^2-a^2}|01\rangle\\\nonumber &&+a\gamma\sqrt
{d^2-c^2}|10\rangle)_{24}|1\rangle_a].\\
\end{eqnarray}

 Finally,
Bob performs a measurement on auxiliary particle $a$ in the basis
$\{|0\rangle_a, |1\rangle_a\}$. If the result of his measurement
 is $|1\rangle_a$, then  the remote  preparation of the original
state fails. If the measurement outcome $|0\rangle_a$ occurs with
probability $\frac
{|ac|^2}{|bd\alpha|^2+|bc\beta|^2+|ad\gamma|^2+|ac\delta|^2}$, then
the remote preparation of a two-particle state $|\phi\rangle$ is
successfully realized.  Evidently, when actual measurement outcome
$|\varphi_\perp\rangle_{13}$ is obtained, the probability of
successfully remote state preparation is
$(|bd\alpha|^2+|bc\beta|^2+|ad\gamma|^2+|ac\delta|^2)\frac
{|ac|^2}{|bd\alpha|^2+|bc\beta|^2+|ad\gamma|^2+|ac\delta|^2}=|ac|^2$.

Similarly, by Eq.(12), if Alice's measurement result on particles 1
and 3 is $|\psi_\perp\rangle_{13}$ with the probability
$|ad\alpha|^2+|ac\beta|^2+|bd\gamma|^2+|bc\delta|^2$, the state of
particles 2 and 4 will become
\begin{equation}
|\psi\rangle_{24}=\frac
{(ac\beta|00\rangle-ad\alpha|01\rangle-bc\delta|10\rangle+bd\gamma|11\rangle)_{24}}{\sqrt
{|ad\alpha|^2+|ac\beta|^2+|bd\gamma|^2+|bc\delta|^2}}.
\end{equation}
Now Bob operates the following unitary transformation
\begin{equation}
U'_1=(|0\rangle\langle 0|-|1\rangle\langle
1|)_2\otimes(-|0\rangle\langle 1|+|1\rangle\langle 0|)_4
\end{equation} on particles 2 and 4. Hence  the state shown in
Eq.(18) was transformed into
\begin{equation}
|\psi'\rangle_{24}=\frac
{(ad\alpha|00\rangle+ac\beta|01\rangle+bd\gamma|10\rangle+bc\delta|11\rangle)_{24}}{\sqrt
{|ad\alpha|^2+|ac\beta|^2+|bd\gamma|^2+|bc\delta|^2}}.
\end{equation}
Next, Bob introduces an auxiliary particle $a$ with the initial
state $|0\rangle_a$ and performs a unitary transformation
\begin{eqnarray}
U'_2=\left (\begin{array}{cccccccc} \frac {c}{d}&C&0&0& &0&0&0\\
C& -\frac {c}{d}&0&0&0&0&0&0\\
0&0&1&0&0&0&0&0\\
0&0&0&-1&0&0&0&0\\
0&0&0&0&\frac
{ac}{bd}&A&0&0\\
0&0&0&0&A&-\frac
{ac}{bd}&0&0\\
0&0&0&0&0&0&\frac {a}{b}&B\\
0&0&0&0&0&0&B&-\frac {a}{b}\\
\end{array}\right)
\end{eqnarray}
on particles 2, 4 and $a$ under the basis $\{ |000\rangle_{24a},$
$|001\rangle_{24a},$ $|010\rangle_{24a},|011\rangle_{24a},$ $
|100\rangle_{24a},$ $|101\rangle_{24a},$ $|110\rangle_{24a},$
$|111\rangle_{24a}\}$. It is straightforward to verify that the
resulting state of of particles 2, 4 and $a$ is
\begin{eqnarray}
\nonumber|\psi''\rangle_{24a} &=&\frac {1}{\sqrt
{|ad\alpha|^2+|ac\beta|^2+|bd\gamma|^2+|bc\delta|^2}}\\\nonumber
&&[ac(\alpha|00\rangle+\beta|01\rangle+\gamma|10\rangle+\delta|11\rangle)_{24}|0\rangle_a\\\nonumber
&&+(a\alpha \sqrt {d^2-c^2}|00\rangle+\gamma\sqrt
{(bd)^2-(ac)^2}|10\rangle\\\nonumber &&+c\delta\sqrt
{b^2-a^2}|11\rangle)_{24}|1\rangle_a].\\
\end{eqnarray}
The above equation shows that Bob can construct a two-particle
state, which Alice wishes to prepare remotely, with certain
probability by performing a measurement on auxiliary particle $a$ in
the basis $\{|0\rangle_a,|1\rangle_a\}$. If Bob's actual measurement
result is $|0\rangle_a$, then remote state preparation is
successful; otherwise remote state preparation  fails. It is easy to
prove that the successful probability of
 remote state preparation in this case is
$(|ad\alpha|^2+|ac\beta|^2+|bd\gamma|^2+|bc\delta|^2)\frac
{|ac|^2}{|ad\alpha|^2+|ac\beta|^2+|bd\gamma|^2+|bc\delta|^2}=|ac|^2
$.

However, if the Alice's actual measurement outcome  on particles 1
and 3 is $|\varphi\rangle_{13}$ ($|\psi\rangle_{13}$), Bob will
obtain the state
$(ac\alpha|00\rangle+ad\beta^*|01\rangle+bc\gamma|10\rangle+bd\delta^*|11\rangle)_{24}$
($(ac\gamma|00\rangle+ad\delta^*|01\rangle-bc\alpha|10\rangle-bd\beta^*|11\rangle)_{24}$).
Since Bob has no knowledge of these states, he can not unitary
convert each of them into the original state, so remote state
preparation fails. But,  when $\alpha$, $\beta$, $\gamma$ and $
\delta$ are real numbers, the situation is not the same. In this
case Bob can prepare the state in every  Alice's measurement outcome
with certain probability. For saving space we omit the concrete
steps for preparing.

Synthesizing two cases (the Alice's the actual measurement outcome
  is either $|\varphi_\perp\rangle_{13}$ or
$|\psi_\perp\rangle_{13}$)
 the  probability of the successful remote
state preparation is $2|ac|^2$.  If $|a|=|b|=|c|=|d|=\frac {1}{\sqrt
2},$ namely,
 the quantum channel consists of two Einstein-Podolsky-Rosen pairs, the  probability equals to 50\%.

In summary, we have presented a scheme of the remote preparation of
a two-particle state  via two Einstein-Podolsky-Rosen pairs and two
partial entangled two-particle state, respectively. The two-particle
state  can be
 prepared probabilistically if the sender performs a
measurement and the receiver introduces an appropriate unitary
transformation with  the aid of long-range Einstein-Podolsky-Rosen
 correlation and transmission of two bits of classical
information. Furthermore we obtained the probability of the successful remote state preparation of the
two-particle state.

{\noindent\bf Acknowledgments}\\[0.2cm]

 This work was supported by the National  Natural Science Foundation of
China under Grant No: 10671054 and Hebei Natural Science Foundation
of China under Grant Nos: A2005000140;07M006, and the Key Project of
Science and Technology Research of Education Ministry of China under
Grant No:207011.

\end{document}